\documentclass[12pt,preprint]{aastex}

\newcommand{\ud}{\mathrm{d}}

\begin{document}


\title{A Very High Energy Gamma-Ray Spectrum of 1ES~2344+514}


\author{M. Schroedter\altaffilmark{1,2}, H. M. Badran\altaffilmark{3}, J. H. Buckley\altaffilmark{4}, J. Bussons Gordo\altaffilmark{5}, D. A. Carter-Lewis\altaffilmark{2}, C. Duke\altaffilmark{6}, D. J. Fegan\altaffilmark{7}, S. F. Fegan\altaffilmark{8}, J. P. Finley\altaffilmark{9}, G. H. Gillanders\altaffilmark{10}, J. Grube\altaffilmark{11}, D. Horan\altaffilmark{1}, G. E. Kenny\altaffilmark{10}, M. Kertzman\altaffilmark{12}, K. Kosack\altaffilmark{4}, F. Krennrich\altaffilmark{2}, D. B. Kieda\altaffilmark{13}, J. Kildea\altaffilmark{14}, M. J. Lang\altaffilmark{9}, Kuen Lee\altaffilmark{4}, P. Moriarty\altaffilmark{15}, J. Quinn\altaffilmark{7}, M. Quinn\altaffilmark{15}, B. Power-Mooney\altaffilmark{7}, G. H. Sembroski\altaffilmark{9}, S. P. Wakely\altaffilmark{16}, V. V. Vassiliev\altaffilmark{8}, T. C. Weekes\altaffilmark{1}, J. Zweerink\altaffilmark{8} }


\altaffiltext{1}{Whipple Observatory, Harvard-Smithsonian Center for Astrophysics, P.O. Box 6369, Amado, AZ 85645-0097}
\altaffiltext{2}{Dept. of Physics and Astronomy, Iowa State University, Ames, IA 50011-3160 }
\altaffiltext{3}{Department of Physics, Faculty of Science, Tanta University, Tanta 31527, Egypt}
\altaffiltext{4}{Department of Physics, Washington University, St. Louis, MO 63130, USA}
\altaffiltext{5}{Instituto de Fisica de Cantabria (CSIC-UC), Avenida de los Castros s/n, 39005 Santander, Spain}
\altaffiltext{6}{Department of Physics, Grinnell College, Grinnell, IA 50112-1690, USA}
\altaffiltext{7}{Experimental Physics Department, National University of Ireland, Belfield, Dublin 4, Ireland}
\altaffiltext{8}{Department of Physics and Astronomy, University of California, Los Angeles, CA 90095-1547, USA}
\altaffiltext{9}{Department of Physics, Purdue University, West Lafayette, IN 47907, USA}
\altaffiltext{10}{Physics Department, National University of Ireland, Galway, Ireland}
\altaffiltext{11}{School of Physics and Astronomy, University of Leeds, Leeds, LS2 9JT, UK}
\altaffiltext{12}{Department of Physics and Astronomy, DePauw University, Greencastle, IN 46135-0037, USA}
\altaffiltext{13}{Physics Department, University of Utah, Salt Lake City, UT 84112, USA}
\altaffiltext{14}{Physics Department, McGill University, Montreal, QC H3A 2T8, Canada}
\altaffiltext{15}{Department of Physical and Life Sciences, Galway-Mayo Institute of Technology, Dublin Road, Galway, Ireland}
\altaffiltext{16}{Enrico Fermi Institute, University of Chicago, Chicago, IL 60637, USA}

%

\begin{abstract}
The BL Lacertae (BL Lac) object 1ES~2344+514 (1ES~2344), at a redshift of 0.044, was discovered as a source of very high energy (VHE) gamma rays by the Whipple Collaboration in 1995 \citep{2344Catanese98}. This detection was recently confirmed by the HEGRA Collaboration \citep{2344Hegra03}. As is typical for high-frequency peaked blazars, the VHE gamma-ray emission is highly variable. On the night of 20 December, 1995, a gamma-ray flare of 5.3-sigma significance was detected, the brightest outburst from this object to-date. The emission region is compatible with a point source. The spectrum between 0.8 TeV and 12.6 TeV can be described by a power law
\begin{displaymath}
\frac{\ud^3 N}{\ud E\ \ud A\ \ud t}=(5.1\pm1.0_{st}\pm1.2_{sy})\times10^{-7} (E/\mathrm{TeV})^{-2.54 \pm0.17_{st}\pm0.07_{sy}}\ \mathrm{\frac{1}{TeV\ m^2\ s}}.
\end{displaymath}
Comparing the spectral index with that of the other five confirmed TeV blazars, the spectrum of 1ES~2344 is similar to 1ES~1959+650, located at almost the same distance. The spectrum of 1ES~2344 is steeper than the brightest flare spectra of Markarian 421 (Mrk~421) and Markarian 501 (Mrk~501), both located at a distance about 2/3 that of 1ES~2344, and harder than the spectra of PKS~2155-304 and H~1426+428, which are located almost three times as far. This trend is consistent with attenuation caused by the infrared extragalactic background radiation.
\end{abstract}





\section{Introduction}
To date, the only confirmed extragalactic gamma-ray sources at energies $>$ 100 GeV (very high energy, VHE) are BL Lacertae (BL Lac) objects and the giant radio galaxy M87 \citep{M87_HESS_Beilicke04}. BL Lac objects are active galactic nuclei (AGN) with (1) characteristic radio/optical/X-ray flux, (2) the absence of emission lines with observed equivalent width greater than 5\AA, and (3) a CA II "break strength" smaller than 25\%  \citep{BLLacsFromEinsteinSS_Perlman96}. These criteria define an object with strong nonthermal emission which almost completely masks the thermal emission from the surrounding host galaxy.  The spectrum, in a $\nu F_{\nu}$ representation, shows a double-peaked structure. The only type of BL Lac objects detected so far to emit VHE emission are high-frequency peaked. For these objects, the low-energy component peaks in the soft to hard X-ray regime and the high-energy component peaks in the VHE regime. The six confirmed VHE BL Lac objects are: Mrk~421~\citep{Mrk421_Punch92,Mrk421_Petry96}, Mrk~501 \citep{Mrk501_Quinn96, Mrk501_Bradbury97}, 1ES~2344+514 \citep{2344Catanese98, 2344Hegra03}, 1ES~1959+650 \citep{1es1959Nishiyama99, 1es1959Holder03, 1es1959Aharonian03}, PKS~2155-304 \citep{pks2155_chadwick99, PKS2155_Aharonian05}, and H~1426+428 \citep{H1426_horan02, HEGRA_H1426_Aharonian02}. The emission level around the two peaks is highly variable, and changes in the spectral shape with flux level have been measured for Mrk~421 \citep{spect_var_Mrk421_Krennrich02,spect_var_Mrk421_Aharonian02, Mrk421_Krennrich03}, Mrk~501 \citep{Mrk501_Djannati99,Mrk501_Aharonian01}, and 1ES~1959+650 \citep{1es1959Aharonian03}. For the other three BL Lacs, variations of the spectral shape with flux level have neither been established nor ruled out.

The VHE observations reported here were carried out by the VERITAS (previously Whipple Gamma Ray) collaboration using an imaging atmospheric Cherenkov telescope \citep{Crab_Nebula_Weekes89}. The telescope, of 10 m diameter, is located on Mt. Hopkins at an altitude of 2320 m above sea level. At the time of observations, the imaging camera consisted of 109 photomultiplier tubes, each viewing 0.259\degr\ of the sky and arranged in a closed-packed hexagonal pattern. The telescope and the data acquisition are described in \citet{detector_Cawley90}.

The organization of this paper is as follows: The status of observations on 1ES~2344 is summarized in Section~\ref{sec:observational_status}. The VHE data and analysis techniques are presented in Section~\ref{sec:flare_data}. This is followed in Section~\ref{sec:simulation} by a description of the gamma-ray simulations necessary for the spectral reconstruction, including estimation of the gamma-ray energy in Sect.~\ref{sec:selection}. The measured VHE spectra are presented in Section~\ref{sec:spectra} and are briefly discussed and summarized in Section~\ref{sec:conclusion}.

\section{Observational Status of 1ES~2344+514\label{sec:observational_status}}
\begin{figure}
\epsscale{.7}
\plotone{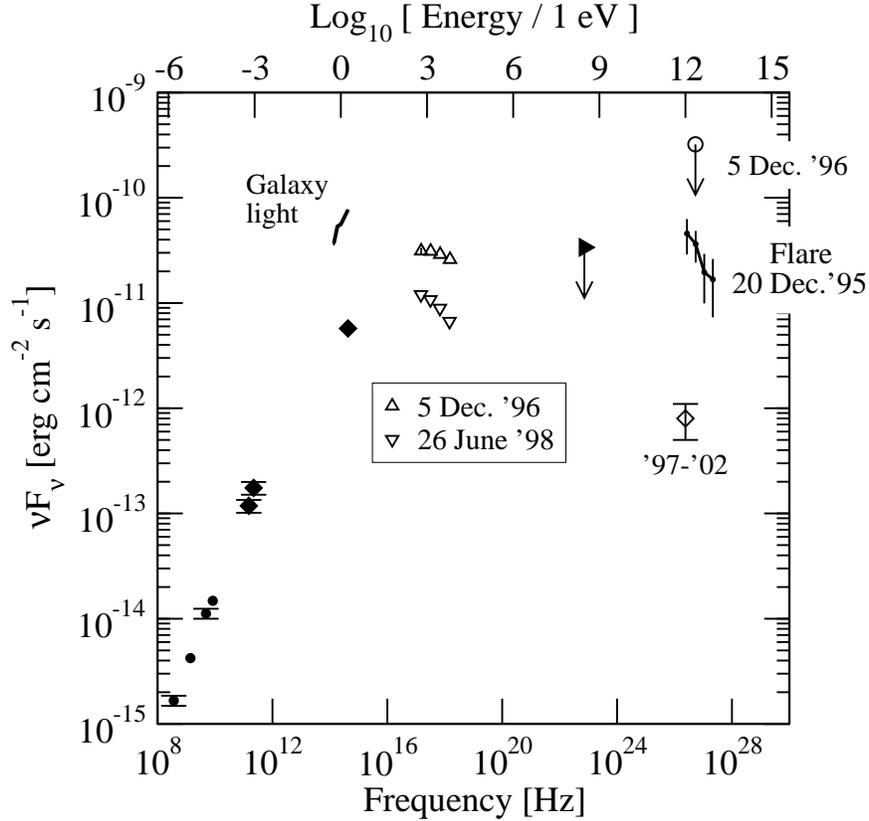}
\caption{The spectral energy distribution of 1ES~2344 along with the VHE flare spectrum obtained with the Whipple 10 m telescope (\emph{line with error bars}). Also shown is the VHE flux upper limit for the night of 5 December, 1996 (\emph{open circle}). Other data were taken from the following sources: 365 MHz from Texas radio survey (\emph{filled circle}) \citep{TexasRadioDouglas1996}, 1.4 GHz from Greenbank  (\emph{filled circle}) \citep{2344Greenbank1.4GHzWhite1992}, 4.85 GHz from Greenbank (\emph{filled circle}) \citep{87GBcatalogGreenbank_4.85GHz1991}, 8.4 GHz from VLA  (\emph{filled circle}) \citep{VLA_8.4GHz_1992}, galaxy photometry at millimeter wavelength (\emph{filled diamond}) \citep{SCUBA1999}, galaxy photometry at K, H, and J-bands from 2MASS (\emph{line segment}) \citep{2MASS_Jarrett03}, galaxy and nucleus R-band photometry obtained with Hubble Space Telescope and corrected for interstellar reddening (\emph{filled diamond}) \citep{HST2000}. X-ray observation with BeppoSAX (\emph{see legend}) from \citep{exceptional_X-ray_var_Giommi_2000}, upper limit at 300 MeV from EGRET (\emph{filled triangle}) \citep{3rd_EGRET_Hartman99}. Quiescent VHE gamma ray flux during the period 1997-2002 from HEGRA (\emph{open diamond}) \citep{2344Hegra03}. }
\label{fig:1es2344_sed}
\end{figure}

The BL Lac object, 1ES~2344+514, at a redshift of 0.044,
was detected in the Einstein Slew Survey \citep{EinsteinSlewSurvey_Elvis92} in the energy range 0.2-4 keV. The survey was constructed from data collected during the HEAO-2 mission from 1978-1981.  1ES~2344 was identified as a BL Lac object in \citet{BLLacsFromEinsteinSS_Perlman96}. The non-contemporaneous spectral energy distribution of 1ES~2344 is shown in Figure~\ref{fig:1es2344_sed}. Observations at all wavelengths show 1ES~2344 to be an unresolved point source. The central black-hole mass is $10^{8.80\pm0.16}$ M$_{\odot}$, derived from stellar velocity dispersion measurements \citep{Barth03_Black_hole_masses}. In the optical regime, a point source with an underlying elliptical host galaxy can be fitted with a radius (half-width at half-maximum) of $r_e=7.12\pm0.02$ kpc ($H_0$=50 km s$^{-1}$ Mpc$^{-1}$ and $q_0$=0) \citep{HST2000}.

%
%
%
%

The optical and far-infrared emission from 1ES~2344 contains significant contributions from the host galaxy. The total photometry by the 2 Micron All Sky Survey \citep{2MASS_Jarrett03} and by the Hubble Space Telescope (HST) \citep{HST2000}, labeled "Galaxy light" in Figure~\ref{fig:1es2344_sed}, lie well above the value expected from pure synchrotron emission in the jet. Observations with the HST in 1996 measured a R-band brightness of the nucleus of 16.83$\pm$0.05 mag from a fit of a point source plus galaxy convolved with the point spread function of the telescope \citep{HST2000}. 
During continued monitoring through 1998, the R-band brightness varied between 16.47 mag \citep{2d_photometric_Nilsson99} and 17.00 mag \citep{optical_imaging_Falomo99}, indicating optical variability. An optical monitoring program in 2000/1 by \citet{photometric_monitoring_Xie2002} found short time scale variability to be weak, with maximum intraday variability of $\Delta V=0.18$ mag, $\Delta R=0.1$ including galaxy light. A relatively large brightness decrease of 0.35 mag was observed in the V-band over 2 weeks in January 2001.

1ES~2344 showed X-ray variability on the time scale of a few hours in the 0.1 - 10 keV energy band during a week-long campaign in 1996 using the BeppoSAX satellite \citep{exceptional_X-ray_var_Giommi_2000}. A follow-up observation in 1998 found 1ES~2344 to be in a very low state, implying a frequency shift by a factor of 30 or more of the peak synchrotron emission. They suggested the interpretation that two distinct electron populations contribute to the synchrotron emission; one steady low-energy component, the other producing soft to hard X-rays with rapid time variability. 

1ES~2344  has been monitored by the Whipple Collaboration since 1995 \citep{2344Catanese98}. Recently, the HEGRA Collaboration reported an independent confirmation of this source \citep{2344Hegra03}. On the night of 5 December, 1996, Whipple VHE and BeppoSAX X-ray observations overlapped for 28 minutes, for which we show the X-ray spectrum and VHE flux upper limit in Figure~\ref{fig:1es2344_sed}. The 99.9\% VHE flux upper limit at energies $>$ 350 GeV was calculated as in \citet{2344Catanese98}.

In the VHE band, the object was observed in a flaring state during the night of 20 December, 1995, with a significance of 5.3 $\sigma$, the strongest gamma-ray flare measured from this object to date. The quiescent flux level of 1ES~2344, compared to the flare presented here, is about 50 times lower \citep{2344Hegra03}. The detection of VHE gamma rays from 1ES~2344 in December 1995, reported by the Whipple Collaboration at the 1997 International Cosmic Ray Conference \citep{2344Catanese1997ICRC}, was considered tentative because follow-up observations by this and other VHE observatories through 1997 did not detect further evidence for gamma-ray emission.  Monitoring by the Whipple Collaboration from 1998 to 2000, however, showed again a small positive excess \citep{2344Badran01}. A summary of all published VHE observations of this source is given in Table~\ref{tab:2344_detections}. An initial measurement of the VHE gamma-ray spectrum covering the entire 1995/6 observing season yielded a spectrum of $(1.14\pm0.50)\times10^{-7}E^{-2.29\pm0.43}$ TeV$^{-1}$ m$^{-2}$ s$^{-1}$, statistical error only, over the energy range $0.5<E<5.0$~TeV with $\chi^2/ndf=3.2/2$ \citep{thesisBussons98, 2344Bussons98}. 

\begin{figure}
\epsscale{.7}
\plotone{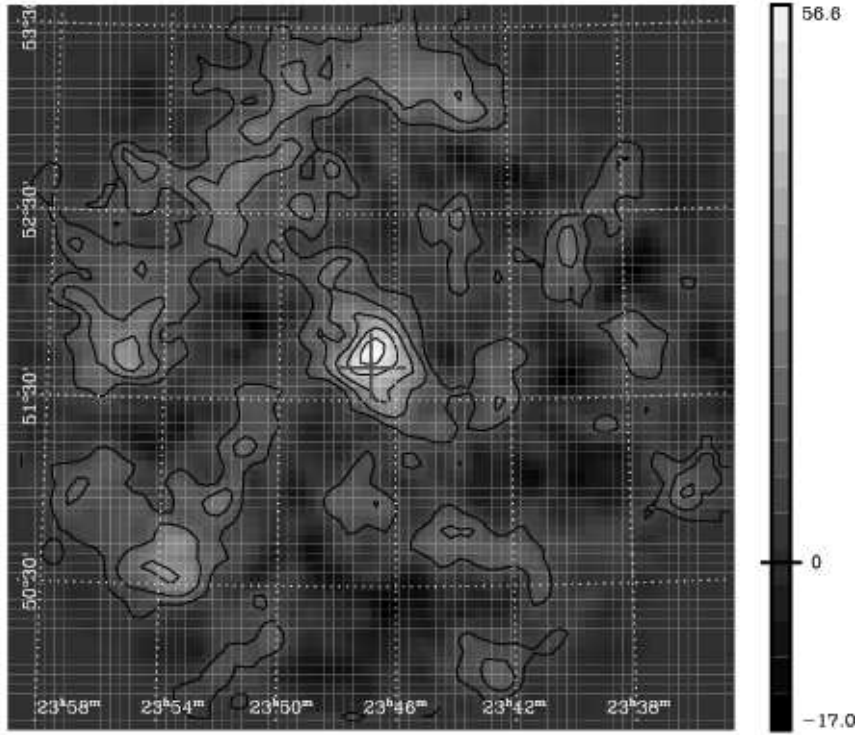}
\caption{Gamma-ray sky map of the field of view around 1ES~2344, with position indicated by the \emph{red cross}. The colors show excess counts with overlaid significance contours in steps of one standard deviation per contour. The \emph{dotted} lines show the RA and DEC.}
\label{fig:2344_2dmap}
\end{figure}

In Figure~\ref{fig:2344_2dmap}, we show the two-dimensional gamma-ray sky map during the flare.The gamma-ray map was constructed from a partial data set, referred to as 'B' in Section~\ref{sec:simulation}.  The emission region is compatible with a point source. This was determined using a Monte-Carlo simulation of the telescope response to a point source of gamma rays. A small telescope pointing error of less than 0.05\degr\ may have been present during the observations, but due to the lack of bright stars in the field of view, we are not able to determine this in retrospect. The centroid of the measured gamma-ray emission is displaced from the known location of 1ES~2344 by RA $0.02\pm0.02$\degr\ and DEC $0.03\pm0.02$\degr. With in a conservatively estimated 0.1\degr\ circle of confusion are located three galaxies and two radio sources, but no other X-ray sources. Thus, the gamma-ray emission likely originated from 1ES~2344.


The EGRET 95\% confidence level upper limit for 1ES~2344 is $6.98\times10^{-8}$ cts cm$^{-2}$ s$^{-1}$, E $>$ 100 MeV \citep{3rd_EGRET_Hartman99}. The peak response for most sources detected with EGRET lies at around 300 MeV, this corresponds to an upper limit at 300 MeV of about $3.4\times10^{-11}$erg cm$^{-2}$.

%
%
%
%
\section{\label{sec:flare_data}Gamma-Ray Flare and Background Data}
\begin{figure}
\epsscale{.6}
\plotone{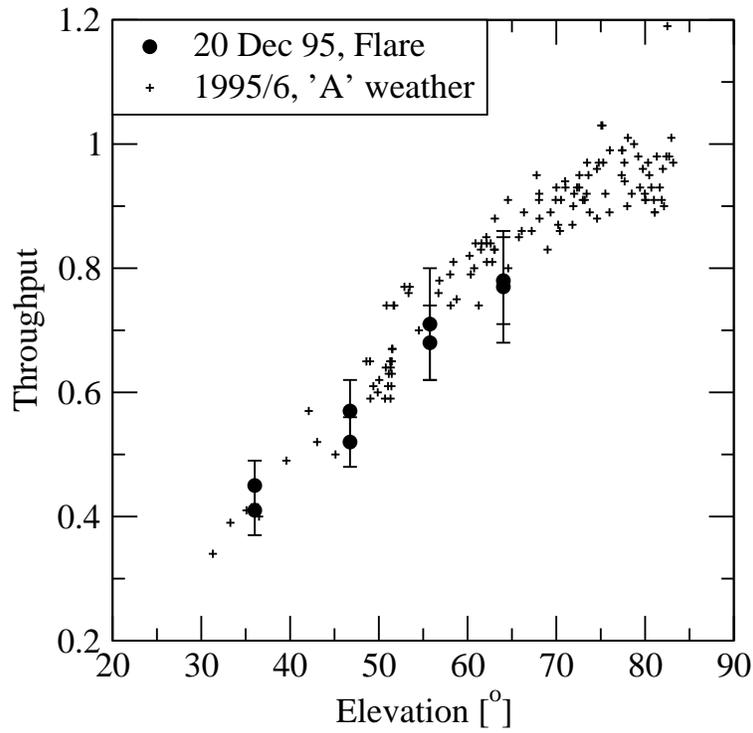}
\caption{Relative cosmic ray rate of 1ES~2344 flare data compared to observations carried out under clear skies (rated ``A'' by observers) in 1995/6. For clarity, the error bars are only shown for the flare data. }
\label{fig:1es2344_throughput}
\end{figure}

Observations with the 10 m telescope were carried out in two pointing modes: (1) with the source in the center of the field of view (ON observation) and (2) with the telescope pointing offset from the source direction by 30 minutes in RA, called OFF observation. The OFF observation is a measurement of the background caused by cosmic rays. On the night of 20 December, 1995, four ON observations were taken with a combined exposure time was 110 minutes. The last ON observation during the night was not complemented with an OFF observation, as is necessary for spectral measurements. Therefore, an OFF observation was selected from the 1995/6 season based on its similarity to the ON observation in elevation, cosmic-ray rate, and night sky brightness. For each ON observations, Table~\ref{tab:2344_data} lists the UTC start time, the average observing elevation, the throughput factor for both, ON and OFF observations, and the measured gamma-ray rate. The throughput factor measures the cosmic-ray rate relative to a reference observation taken under clear skies~\citep{throughput_Lebohec03}. The weather during these observations was rated ''A'' by the observers, meaning clear skies. Figure~\ref{fig:1es2344_throughput} confirms this by comparing these observations with other ''A'' weather observations taken between October 1995 and April 1996.
%

The standard analysis method for data taken with the 10 m telescope \citep{Survey1988-91_Reynolds93} includes conditioning of the images, parameterization, and selection of gamma-ray like events. Conditioning of the images consists of: (1) flat-fielding of the relative gain between pixels, (2) equalizing the sky brightness between the ON and OFF observation \citep{padding_Cawley93}, and (3) removing pixels that are below a certain signal-to-noise ratio.  Images are then parameterized by their RMS $width$ and $length$, their $distance$ from the center of the field of view \citep{HillasParameters_Hillas1985}, and the orientation angle of their major axis relative to the pointing direction of the telescope, $alpha$ \citep{HERCULES_Weekes87}. The total amount of light collected is referred to as the $size$ of the image.

The gamma-ray signal is derived from the excess number of events between ON and OFF runs, where only those images are selected that are likely to have been produced by a gamma-ray source located at the center of the field of view. The rate given in Table \ref{tab:2344_data} shows the gamma-ray rate after application of one particular set of selection criteria (cuts), called $Supercuts1995$ \citep{2344Catanese98}. These cuts are not optimal for spectral measurements because the selection efficiency for gamma-rays decreases dramatically with energy. Therefore, a different set of cuts was developed empirically using simulated gamma-ray events, see Section~\ref{sec:selection}. The analysis of the data given in Table~\ref{tab:2344_data} is in agreement with a previous analysis by \citet{2344Catanese98}.

%
%
%
%
\section{\label{sec:simulation}Calibration and Spectral Reconstruction}
The data contain a relatively low gamma-ray rate and were taken over a wide range of elevations. To obtain an accurate energy calibration our analysis technique requires us to analyze different elevation ranges separately. To maintain a good signal-to-noise ratio, the data were combined at the two average elevations of 58\degr\ and 41\degr\ and referred to as datasets A and B, respectively. The data in these two sets were taken sequentially during the night, allowing us to investigate time variability in the emission level.

A total of 500,000 gamma-ray initiated showers were simulated at each elevation; these were distributed in energy randomly according to a power law of index -2.5 and covering a circular area around the telescope axis. The Monte-Carlo simulations of gamma-ray initiated particle showers in the atmosphere and subsequent detection of Cherenkov photons by the telescope were carried out with the Grinnell-ISU (GrISU) package%
\footnote{Available at \url{http://www.physics.utah.edu/gammaray/GrISU/} }%
. At 58\degr\ elevation, simulations were carried out over the energy range 0.1-100 TeV and impact radius less than 300 m. At 41\degr\ elevation, simulations were carried out over the energy range 0.3-100 TeV and impact radius less than 350 m. The low-energy cut-off was chosen to extend beyond the range of energies of events that trigger the telescope. The night-sky brightness level of simulated showers was matched to that measured from the data.

The absolute light throughput of the telescope was calibrated with Cherenkov images of muons recorded by the telescope. For this, only complete muon rings were selected using a specially developed algorithm~\citep{thesis_Schroedter04}. This ensures that the total amount of light is well known. The light throughput factor derived in this way was used to measure the spectrum of the Crab Nebula during 1995/6 season. With statistical (st) and systematic (sy) errors, the spectrum of the Crab Nebula between 0.3 TeV and 13 TeV can be fitted by $(4.2\pm 0.3_{st}\pm0.7_{sy})\times10^{-7} E^{-2.38\pm0.08_{st}\pm0.04_{sy}}\ \mathrm{TeV^{-1}\ m^{-2} s^{-1}}$ with $\chi^2_{min}$/ndf = 3.2/(9-2). This is compatible with other measurements \citep{Mohanty98, CrabHillas98}.

The energy resolution of the spectral analysis depends on rejecting cosmic-ray images, and selecting only those gamma-ray images with well defined image parameters. The collection area near the triggering threshold is difficult to model in the simulations and hence a software cut on the minimum brightness is applied that lies substantially above the hardware threshold. The following set of loose cuts were then applied to data and simulations: 0.31\degr$<distance<$1.1\degr, $length/size<$0.00085 \degr/dc, $max2>65$ dc, and $alpha<25$\degr. 

\begin{figure}
\epsscale{.8}
\includegraphics[width=3in,clip,viewport=0 0 280 287]{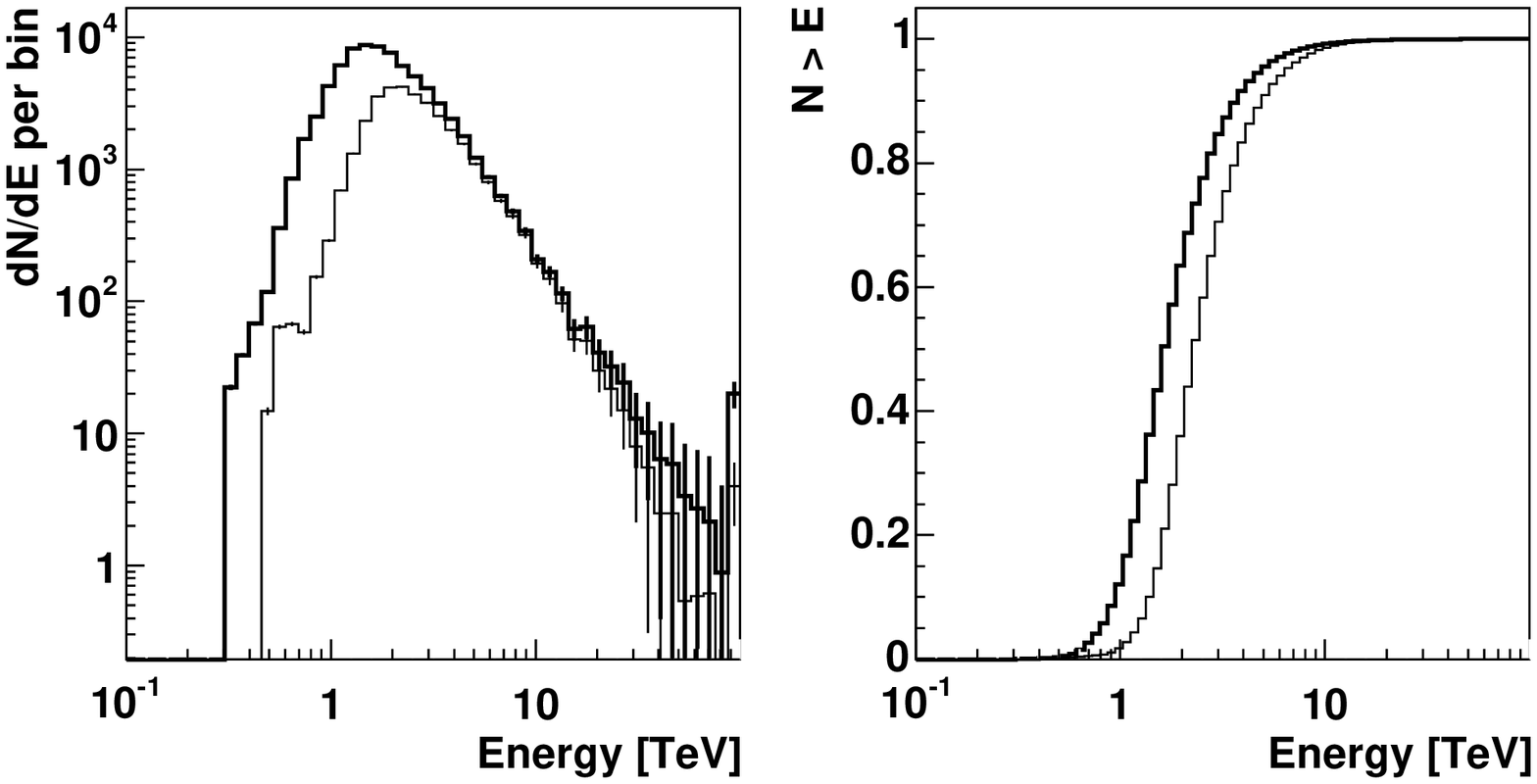}
\includegraphics[width=3in,clip,viewport=0 0 280 287]{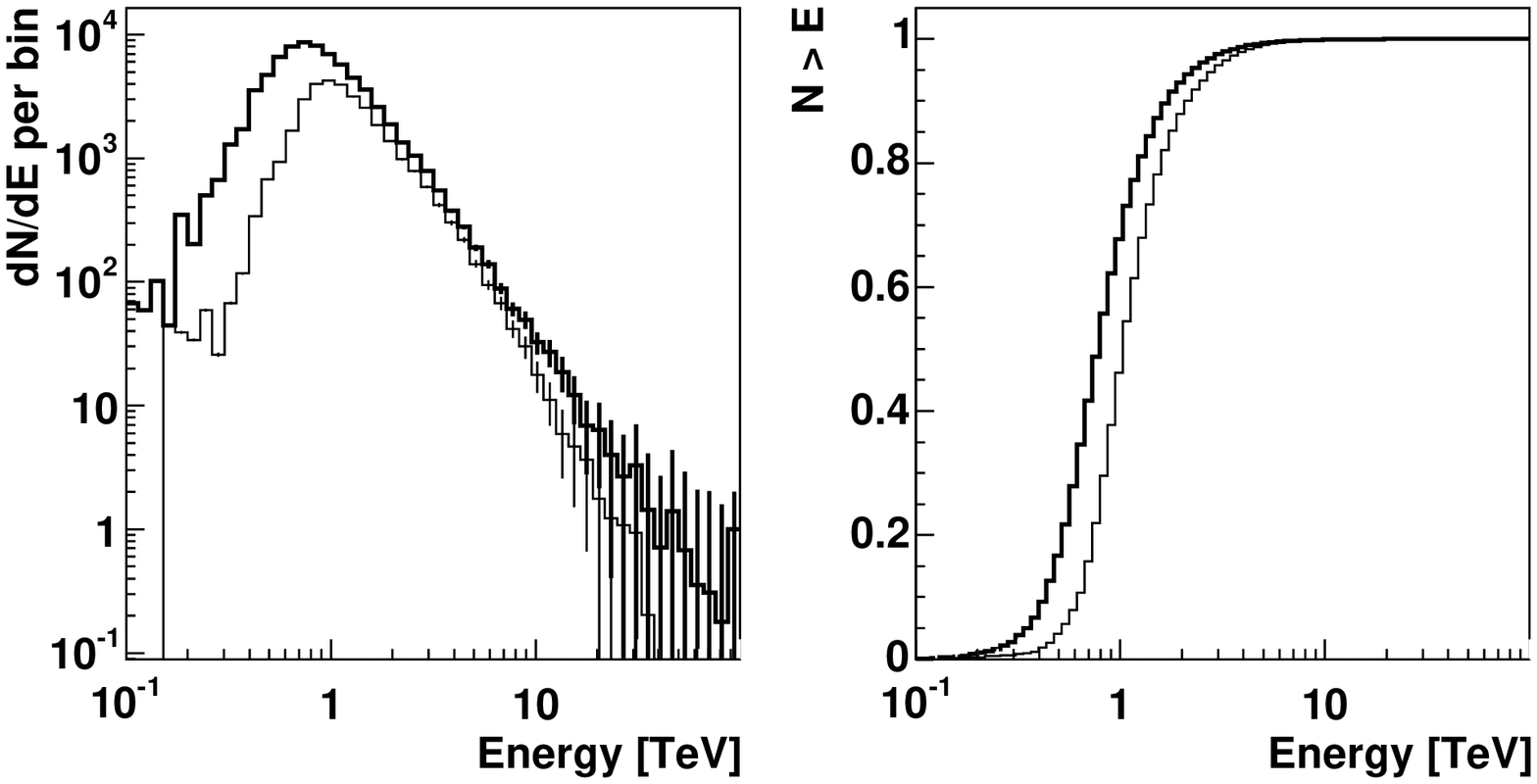}
\caption{Simulated trigger rate at 41\degr\ elevation (\emph{left}) and 58\degr\ (\emph{right}) of gamma-rays distributed with a power law index of -2.5. The two lines show the rate after application of spectral cuts (\emph{bold}) and $Supercuts1995$ (\emph{thin}).}
\label{fig:trigger_1es2344}
\end{figure}

\begin{figure}
\epsscale{1.}
\plottwo{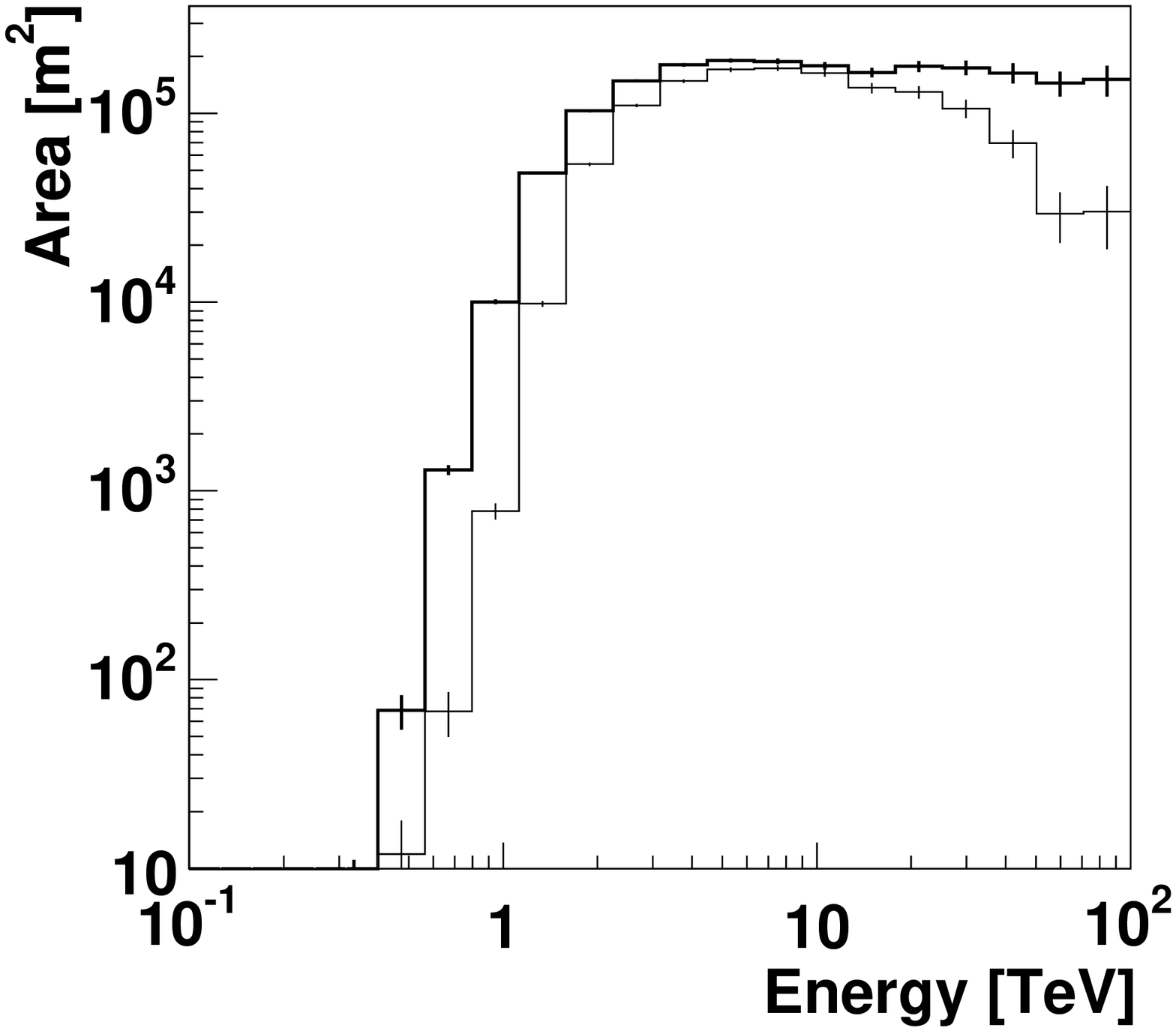}{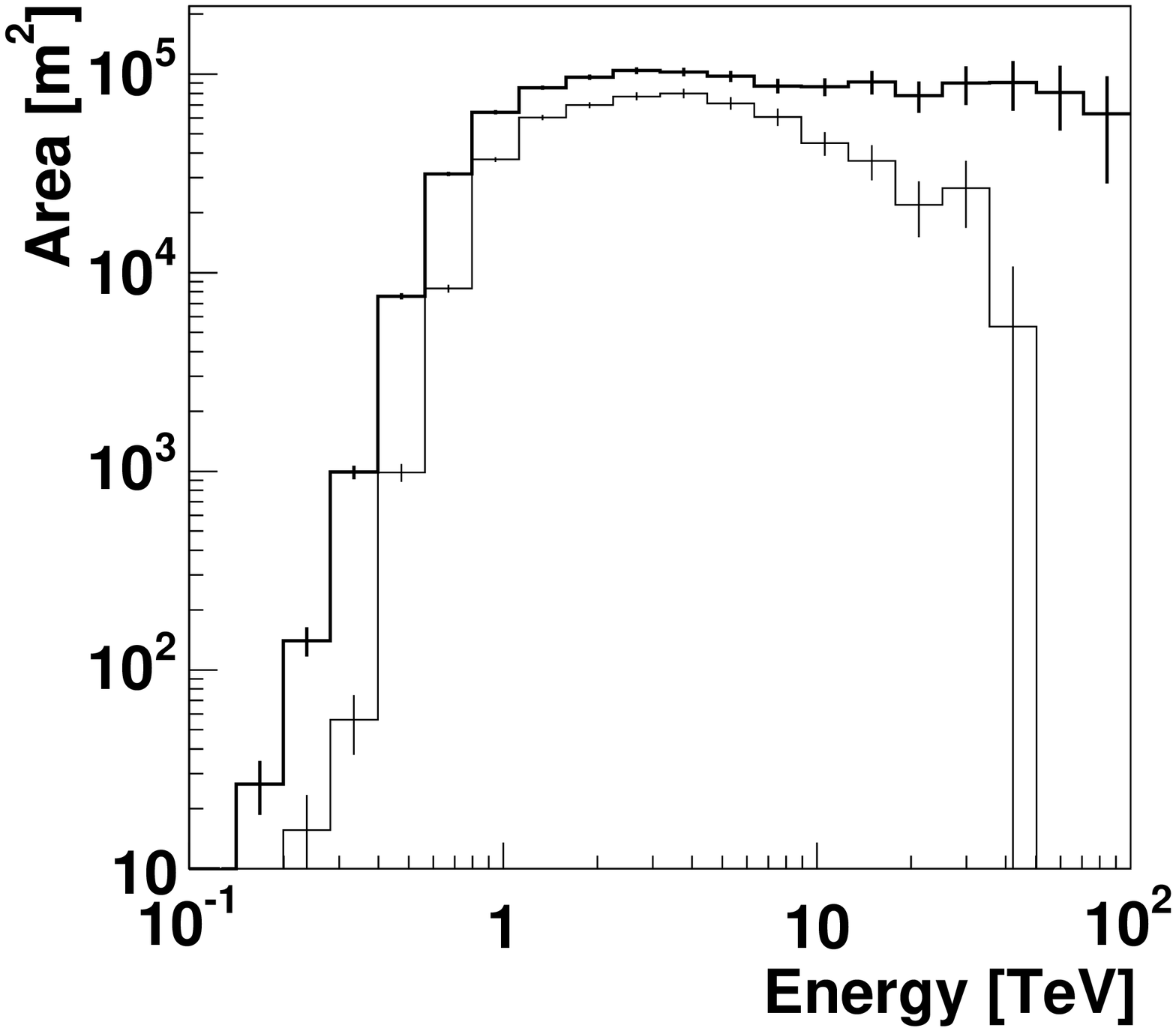}
\caption{Collection area for gamma rays at 41\degr\ elevation (\emph{left}) and at 58\degr\ elevation (\emph{right}) in 1995 for spectral cuts (bold line) and $Supercuts1995$ cuts (thin line).}
\label{fig:area_1es2344}
\end{figure}

The differential trigger rates at 41\degr\ elevation, data set B, are shown in Figure~\ref{fig:trigger_1es2344} for a spectrum with differential index of -2.5. The peak trigger rate occurs at an energy of 1.4 TeV for spectral cuts, described below, and 2.1 TeV with $Supercuts1995$. With these cuts, 90\% of the triggers occur above 1.05 TeV and 1.67 TeV, respectively. The collection area, shown in Figure~\ref{fig:area_1es2344}, reaches 10\% of its maximum value of ~170,000 m$^2$ at an energy of about 1.1 TeV for spectral cuts. The differential trigger rates at 58\degr\ elevation, corresponding to data set A, are shown in Figure~\ref{fig:trigger_1es2344}. The peak trigger rate occurs at an energy of 0.69 TeV for spectral cuts and 1.1 TeV with $Supercuts1995$. For these two sets of cuts, 90\% of the triggers occur above 0.48 TeV and 0.75 TeV, respectively. The collection area, shown in Figure~\ref{fig:area_1es2344}, reaches 10\% of its maximum value of ~136,000 m$^2$ at an energy of about 0.51 TeV for spectral cuts.
%
%
%
%
%
\subsection{\label{sec:selection}Event Selection and Energy Estimation}
The spectral analysis method has been described in \citet{spectrum_H1426_Petry02, Mohanty98}. Simulations at 41\degr\ elevation show an energy resolution of $rms(\Delta \log E)=0.15$ or $rms(\Delta E/E)=0.40$, with an energy estimation bias of $| \Delta \log E| = 0.018$ over the energy region $E$ = 0.8 TeV to 40 TeV. This energy range begins at 10\% of the peak collection area. A cut-off at the high energies is necessary as the limited field of view of the camera truncates large showers and the estimated and true energies begin to diverge. At 58\degr\ elevation, the energy resolution is $rms(\Delta \log E)=0.16$ or $rms(\Delta E/E)=0.49$) and $| \Delta \log E| = 0.012$ over the energy range $E$ = 0.4 TeV to 25 TeV.

\begin{figure}
\epsscale{1.}
\plotone{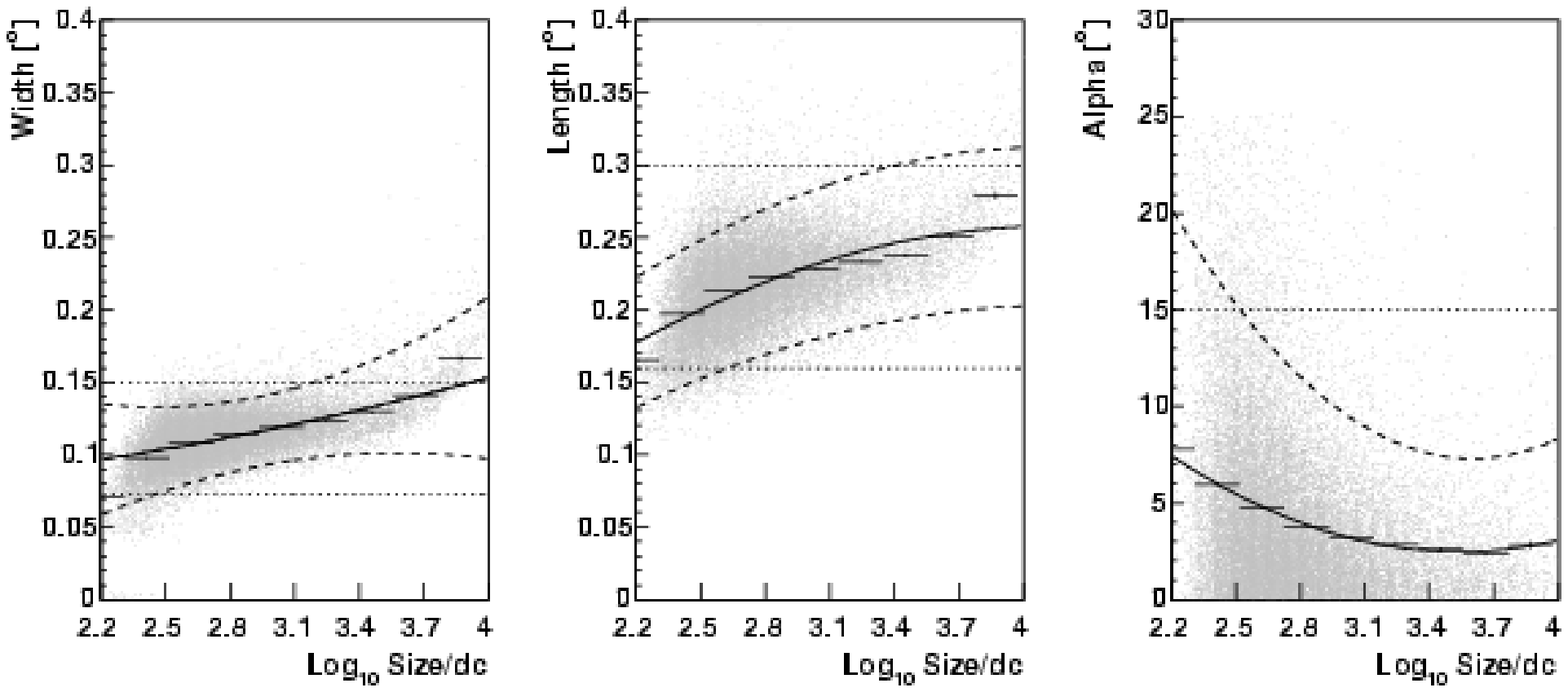}
\plotone{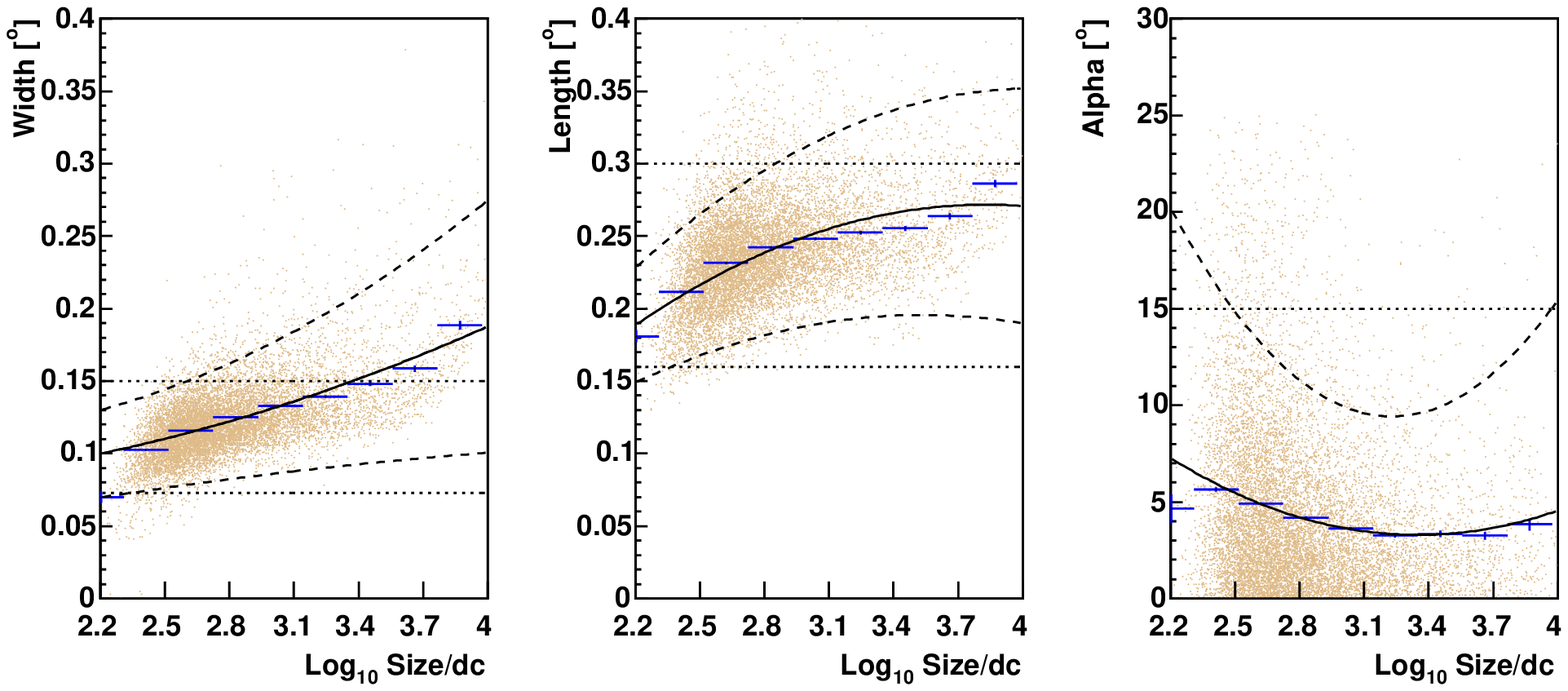}
\caption{Simulated parameter distributions and cut levels with $\log(size)$ after application of loose cuts, see text. \emph{Top row}: simulated at 41\degr\ elevation, \emph{bottom row}: 58\degr\ elevation. The \emph{dots} are simulated events and \emph{crosses} represent the average. The \emph{solid lines} shows the fitted polynomial to the average. \emph{Dashed lines} show the actual cut chosen at a tolerance of two standard deviations around the average. \emph{Dotted lines} show the cut level of $Supercuts1995$.}
\label{fig:cuts_1es2344}
\end{figure}

The gamma-ray signal is contaminated by a large fraction of cosmic-ray events. To reject this background, cuts are imposed on the parameters $distance$, $width$, $length$, and $alpha$. The cuts derived from the Monte-Carlo simulations scale with $size$ so that the efficiency of selecting gamma rays remains unchanged as a function of energy. The fraction of gamma rays passing the cuts for simulations at 41\degr\ elevation is 86\%, and it is 87\% at 58\degr\ elevation. The distributions of the parameters $width$, $length$, and $alpha$ are shown for simulated gamma rays in Figure~\ref{fig:cuts_1es2344}. The cuts are chosen at a nominal 2 standard deviations around the mean value. The simulations at 58\degr\ are limited by statistics at high energies, making the cuts somewhat inefficient. In particular, the upturn at large dc value of the $alpha$-cut is unphysical, but the cut-level still remains below the $Supercuts1995$ value of 15\degr. The unphysical upturn is due to the second order polynomial used in fitting the cut-level. For comparison, the level of $Supercuts1995$ is also shown in Figure~\ref{fig:cuts_1es2344}.
%
%
%
%
%
\section{Flare Spectra\label{sec:spectra}}
\begin{figure}
\epsscale{.6}
\plotone{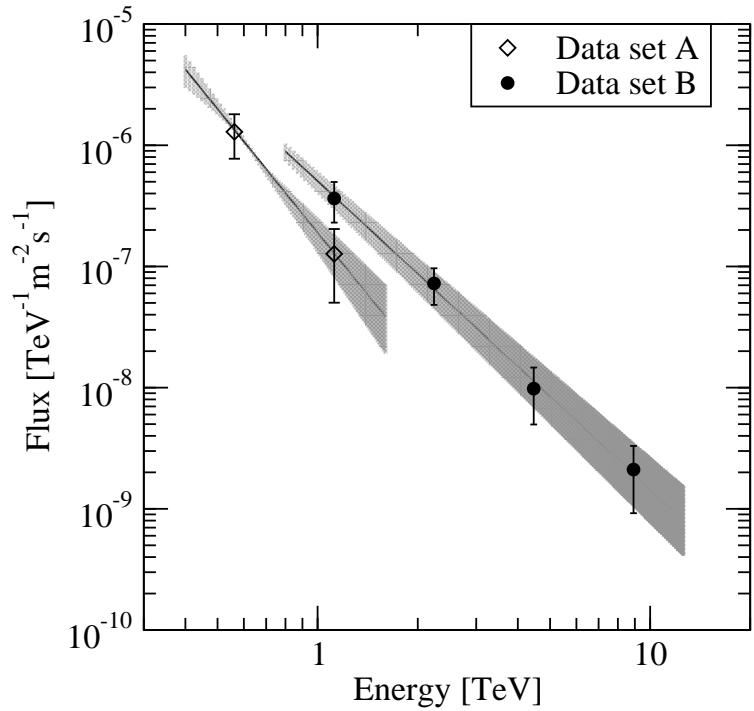}
\caption{Differential flux spectrum of 1ES~2344 on 20 December 1995. Spectra from data sets A (\emph{diamonds}) and B (\emph{circles}) are shown together with power law fits (\emph{solid lines}). The shaded regions show the confidence interval of the power law fits and were obtained by varying both parameters to their individual 68\% confidence interval.}
\label{fig:2344_spectra}
\end{figure}

The number of excess gamma-ray events in each energy bin after application of all cuts is presented for both data sets in Tables ~\ref{tab:2344_events_58deg} and ~\ref{tab:2344_events_41deg}. Due to the very small signal, the bin width is chosen at twice the energy resolution  $\Delta (\log E)=0.3$~\citep{spectrum_H1426_Petry02}. Flux upper limits are given if the gamma-ray significance is less than 1 $\sigma$ in the energy bin. The upper limits are at the 98\% confidence level and calculated according to the method of \citet{upperLimit_Helene83}.
%
%
The spectra for the two data sets A and B are shown in Figure~\ref{fig:2344_spectra}. The error bars show the statistical error only. 
%
%

\begin{figure}
\epsscale{.6}
\plotone{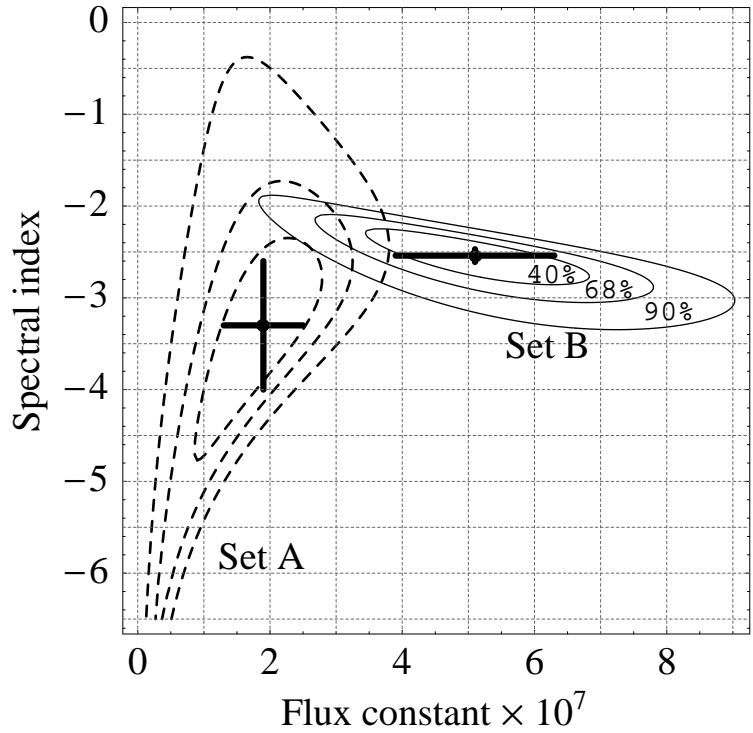}
\caption{Confidence regions corresponding to spectrum of data set A (\emph{dashed lines}) and set B (\emph{solid lines}). Confidence regions are shown with probability content of 40\%, 68\%, and 90\% for the simultaneous values of the spectral index and flux constant.  Also shown are the systematic error on the flux constant and spectral index (\emph{crosses})}
\label{fig:2344_contour}
\end{figure}

For dataset B, the power law fit to the spectrum over the energy range from 0.8 TeV to 12.6 TeV is given by
\begin{equation}
\frac{\ud N}{\ud E\ \ud A\ \ud t}=(5.1\pm1.0_{st}\pm1.2_{sy})\times10^{-7} E^{-2.54 \pm0.17_{st}\pm0.07_{sy}}\ \mathrm{\frac{1}{TeV\ m^2\ s}},
\end{equation}
with $\chi^2_{min}/ndf = .2/(4-2)$. The $\chi^2$ probability for this data to randomly arise from the power-law fit is 0.9. The statistical error represents the 68\% confidence interval (CI) for a fit with one free parameter while the other parameter is frozen at its optimum value. The 68\% CI with two simultaneous free parameters, defined by $\chi_{min}^2+2.3$, is shown in Figure~\ref{fig:2344_contour}. 

The systematic errors of the flux constant and spectral index arising from the energy calibration and the cut-tolerance are indicated in Figure~\ref{fig:2344_contour} by crosses. The cut tolerance, with a nominal value of 2 standard deviations, was varied between 1.5 and 2.5 standard deviations; the level of the muon-based energy calibration is $\pm$10\%. The uncertainty in the energy calibration affects mostly the flux constant. For example, a 10\% change in the energy calibration changes the flux constant by 25\% (30\%) if the spectrum has a differential index of -2.5 (-3.0). In addition, due to the large elevation range covered, a small systematic uncertainty on the order of 10-15\% is intrinsic to the GrISU simulations \citep{Mrk421Krennrich99}. The spectral index is affected mostly by varying the cut tolerance. It should be noted that the systematic error evaluated in this way is smaller than the statistical error. This means that a good estimate of the systematic error is not possible with this method; nevertheless it does indicate the relative importance of the two sources of error.

For data set A, the power law fit over the energy range from 0.4 TeV to 1.6 TeV is given by
\begin{equation}
\frac{\ud N}{\ud E\ \ud A\ \ud t}=(1.9\pm0.6_{st}\pm0.6_{sy})\times10^{-7} E^{-3.3 \pm0.7_{st}\pm0.7_{sy}}\ \mathrm{\frac{1}{TeV\ m^2\ s}},
\end{equation}
and the confidence interval contours are shown in Figure~\ref{fig:2344_contour}. 

As the spectral indexes of the two spectra are compatible, it is possible to adjust the flux constant of the less significant spectrum (set A) so that it overlaps, in a least-squares sense, with the spectrum of set B. However, as the statistical significance of data set A is very small compared to set B, combining the two data sets results in an insignificant improvement in the statistical error of the spectral index. Therefore, the spectral measurement of 1ES~2344 derived here, is best represented by spectrum of data set B, alone.

\section{\label{sec:conclusion}Discussion}
1ES~2344 is a variable source; during the flare on 20 December, 1995, the gamma-ray emission from 1ES~2344 was about 50 times brighter than during the quiescent phase measured several years later. To obtain an accurate energy calibration, our analysis technique required us to analyze data taken at different observing elevations separately. Therefore, we split the data into two sets, A and B, with 56 and 38 minutes exposure time, respectively. The data sets were taken consecutively during the night. The spectral indices measured from the two data sets are compatible with each other. The increase of the flux constant over the two hours of observation, though also not very significant, is not unexpected as very large variability of the VHE flux on time scale of hours has been observed for other blazars~\citep{Mrk421_Gaidos96,Mrk501_Quinn96,1es1959Holder03,PKS2155_Aharonian05}.

The measured VHE spectra are attenuated through pair production with the infrared extragalactic background light (EBL) \citep{AbsorptionNikishov62}. Due to the EBL spectral shape, the attenuation manifests itself as a steepening of the measured VHE spectrum between roughly 1 and 5 TeV and becomes more pronounced with larger redshift. A cut-off feature is thus expected in the VHE spectra, if variations in the intrinsic VHE spectrum are ignored. Such a cut-off feature has been established for Mrk~421 and Mrk~501 \citep{Mrk501_Aharonian01, Mrk421_Krennrich01, spect_var_Mrk421_Krennrich02}. Their spectra can be described with a power law with exponential cut-off: $dN/dE \propto E^{-\alpha} \exp{(E/E_0)}$. The cut-off energy, $E_0$, differs between the two blazars by 2.6$\pm$1.2 TeV \citep{spect_var_Mrk421_Aharonian02}. Unfortunately, for 1ES~2344 the low statistical significance of the spectrum precludes the measurement of such a cut-off energy.

VHE spectra are now available for all six confirmed TeV blazars. The power law spectral indexes of fits to the brightest flares from the blazars appear to steepen with redshift \citep{EBL_Schroedter05}. The spectral index of the 1ES~2344 VHE flare is steeper than the brightest flare spectra of Mrk~421 and Mrk~501, both located at about 2/3 the distance of 1ES~2344. The flare spectra of PKS~2155-304 and H~1426+428, located almost three times as far, are softer than that of 1ES~2344. The spectral index of the 1ES~2344 flare is similar to the flare spectrum of 1ES~1959+650, which is located at almost the same redshift. This trend is consistent with attenuation caused by the infrared extragalactic background radiation \citep{EBL_Schroedter05, Steepening_Stecker99}. Alternatively, galaxy evolution might be responsible for the observed spectral steepening with redshift. For example, if younger galaxies have enhanced mid-infrared radiation nearer to the central black hole, then this would produce gamma-ray attenuation indistinguishable from that caused by the EBL.

No contemporaneous measurements at other wavelengths were taken during the gamma-ray flare of 1ES~2344 on 20 December, 1995. This precludes the application of models to constrain the gamma-ray production mechanism, because the gamma-ray emission is known to be highly variable. Almost one year later, on 5 December, 1996, a simultaneous TeV / X-ray observation occurred together with the \emph{BeppoSAX} satellite. The detailed X-ray spectrum measured during this night \citep{exceptional_X-ray_var_Giommi_2000} is complemented, however, only by an upper limit of the TeV flux, again precluding models to be significantly constrained.

%




\acknowledgments
We acknowledge the technical assistance of K. Harris, T. Lappin, and E. Roache. This research has made use of the NASA/IPAC Extragalactic Database (NED) which is operated by the Jet Propulsion Laboratory, California Institute of Technology, under contract with the National Aeronautics and Space Administration.

\clearpage



\clearpage

\clearpage

\clearpage


\clearpage

\clearpage

\clearpage

\clearpage

\clearpage

\begin{table}\begin{center}
\caption{Summary of VHE measurements of 1ES~2344.}
\label{tab:2344_detections}
\begin{tabular}{|c|c|c|c|c|c|}
\tableline\tableline
Date        &Reference &Exposure&S\tablenotemark{a}&Integral Flux&$E_{thresh}$\\
            &          &[hr]&[$\sigma$]&[$\times10^{-7}$ m $^{-2}$s$^{-1}$] &[TeV]\\ \tableline
1995/6      & \citealt{2344Catanese98}&20.5 &5.8 &$1.7\pm0.5$&0.35\\
20 Dec. 1995& \citealt{2344Catanese98}&1.85 &5.3\tablenotemark{b}&$6.6\pm1.9$&0.35\\
1996/7      & \citealt{2344Catanese98}&24.9 &0.4 &$<0.82$\tablenotemark{c}&0.35\\
Dec. 1997   & \citealt{2344Hegra2000} &15.8 &NA &$<0.29$\tablenotemark{d}&1.0\\
1997-2002   & \citealt{2344Hegra03}   &72.5 &4.4 &$0.08\pm0.03$&0.8\\
1998        & \citealt{2344Hegra99}   &23.8 &3.3\tablenotemark{b} &$<0.09$\tablenotemark{d}  &1.0\\
2000        & \citealt{2344Badran01}  &3.1  &2.4&$1.1\pm0.1$\tablenotemark{e}&$\approx$0.4\\
\tableline
\end{tabular}
\tablenotetext{a}{Statistical excess.}
\tablenotetext{b}{Part of the data listed in the above entry.}
\tablenotetext{c}{99.9\% C.L. upper limit}
\tablenotetext{d}{99\% C.L. upper limit}
\tablenotetext{e}{Statistical error only.}
\end{center}\end{table}

\clearpage

\begin{table}\begin{center}
\caption{Details of the observations taken on 20 Dec. 1995.}
\label{tab:2344_data}
\begin{tabular}{|r|r|r|r|}
\tableline\tableline
UTC\tablenotemark{a}&El.&Throughput\tablenotemark{b}&$\gamma$-rate\tablenotemark{c}\\
\tableline
2:32 &64$^{\circ}$&0.77/0.78 $\pm$0.08&$0.70 \pm 0.28$\\
3:34 &55$^{\circ}$&0.68/0.71 $\pm$0.08&$1.04 \pm 0.37$\\
4:35 &47$^{\circ}$&0.52/0.57 $\pm$0.05&$0.91 \pm 0.42$\\
5:36 &37$^{\circ}$/36$^{\circ}$&0.45/0.41 $\pm$0.04&$1.54 \pm 0.47$\\
\tableline
\multicolumn{3}{|r|}{Average}                      &$1.14 \pm0.20$\\  
\tableline
\end{tabular}
\tablenotetext{a}{Start time of the ON observation. The first three ON observations lasted for 28 minutes followed by an OFF observation. The last observation had a length of only 10 min. A complementary OFF observation, necessary for the spectral analysis, was chosen based upon similar observation conditions. }
\tablenotetext{b}{Relative cosmic-ray rate for ON/OFF observation, see text.}
\tablenotetext{c}{Gamma-ray rate per minute after $Supercuts1995$ \citep{2344Catanese98}.}
\end{center}\end{table}

\clearpage

\begin{table}\begin{center}
\caption{Event statistics and flux in each energy bin for data set A. Upper limits are given at the 98\% confidence level.}
\label{tab:2344_events_58deg}
\begin{tabular}{|c|c|c|r @{$\pm$} l |c|r @{$\pm$} l |}
\tableline\tableline
Energy & ON & OFF & \multicolumn{2}{|c|}{ON-OFF} & S & \multicolumn{2}{|c|}{Flux} \\ 

[TeV] & [events] &  [events] & \multicolumn{2}{|c|}{[events]} & [$\sigma$] & \multicolumn{2}{|c|}{[TeV$^{-1}$m$^{-2}$s$^{-1}$]}  \\ 
\tableline
 0.56 & 63 & 38 & 25 &  10 &  2.5 & (1.29 & 0.51)$\times10^{-6}$\\ 
 1.12 & 83 & 63 & 20 &  12 &  1.7 & (1.27 & 0.77)$\times10^{-7}$\\ 
 2.24 & 39 & 42 & -3 &   9 & -0.3 & \multicolumn{2}{|c|}{$<$3.91$\times10^{-8}$}\\ 
 4.47 & 22 & 19 & 3 &   6 &  0.5  & \multicolumn{2}{|c|}{$<$1.46$\times10^{-8}$ }\\ 
 8.91 & 8 & 7 & 1 &   4 &  0.3    & \multicolumn{2}{|c|}{$<$3.62$\times10^{-9}$}\\ 
\tableline
Total & 220 & 174 & 44 & 19.8 & 2.2&\multicolumn{2}{|c|}{}\\ 
\tableline
\end{tabular}
\end{center}\end{table}

\clearpage

\begin{table}\begin{center}
\caption{Event statistics and flux in each energy bin for data set B. Upper limits are given at the 98\% confidence level.}
\label{tab:2344_events_41deg}
\begin{tabular}{|c|c|c|r @{$\pm$} l |c|r @{$\pm$} l |}
\tableline\tableline
Energy & ON & OFF & \multicolumn{2}{|c|}{ON-OFF} & S & \multicolumn{2}{|c|}{Flux} \\ 

[TeV] & [events] &  [events] & \multicolumn{2}{|c|}{[events]} & [$\sigma$] & \multicolumn{2}{|c|}{[TeV$^{-1}$m$^{-2}$s$^{-1}$]}  \\ 
\tableline
 1.12 & 55 & 30 & 25 &   9 &  2.7 & (3.64 & 1.34)$\times10^{-7}$\\ 
 2.24 & 86 & 51 & 35 &  12 &  3.0 & (7.24 & 2.42)$\times10^{-8}$\\ 
 4.47 & 35 & 20 & 15 &   7 &  2.0 & (9.82 & 4.85)$\times10^{-9}$\\ 
 8.91 & 14 & 6 & 8 &   4 &  1.8   & (2.11 & 1.19)$\times10^{-9}$\\ 
17.78 & 7 & 4 & 3 &   3 &  0.9    & \multicolumn{2}{|c|}{$<$1.19$\times10^{-9}$} \\ 
35.48 & 0 & 2 & -2 &   1 & -1.4   & \multicolumn{2}{|c|}{$<$1.43$\times10^{-10}$} \\ 
\tableline
Total & 197 & 113 & 84 & 17.6 & 4.8&\multicolumn{2}{|c|}{}\\
\tableline
\end{tabular}
\end{center}\end{table}

\clearpage


\begin{thebibliography}{58}
\expandafter\ifx\csname natexlab\endcsname\relax\def\natexlab#1{#1}\fi
\expandafter\ifx\csname url\endcsname\relax
  \def\url#1{{\tt #1}}\fi

\bibitem[{Aharonian} et~al.(2002{\natexlab{a}}){Aharonian}, {Akhperjanian},
  {Barrio}, {Beilicke}, {Bernl{\" o}hr}, {B{\" o}rst}, {Bojahr}, {Bolz},
  {Contreras}, {Cornils}, {Cortina}, {Denninghoff}, {Fonseca}, {Girma},
  {Gonzalez}, {G{\" o}tting}, {Heinzelmann}, et~al.]{HEGRA_H1426_Aharonian02}
{Aharonian}~F, {Akhperjanian}~A, {Barrio}~J, et~al.
\newblock {TeV gamma rays from the blazar H 1426+428 and the diffuse
  extragalactic background radiation}.
\newblock {\em \aap}, 384:\penalty0 L23--L26, Mar. 2002{\natexlab{a}}.

\bibitem[{Aharonian} et~al.(2001){Aharonian}, {Akhperjanian}, {Barrio},
  {Bernl{\" o}hr}, {B{\" o}rst}, {Bojahr}, {Wittek},
  et~al.]{Mrk501_Aharonian01}
{Aharonian}~F, {Akhperjanian}~A, {Barrio}~J, et~al.
\newblock {The TEV Energy Spectrum of Markarian 501 Measured with the
  Stereoscopic Telescope System of HEGRA during 1998 and 1999}.
\newblock {\em \apj}, 546:\penalty0 898--902, Jan. 2001.

\bibitem[{Aharonian} et~al.(2002{\natexlab{b}}){Aharonian}, {Akhperjanian},
  {Beilicke}, {Bernl{\" o}hr}, {B{\" o}rst}, {Bojahr}, {Bolz}, {Coarasa},
  {Contreras}, {Cortina}, {Costamante}, {Denninghoff}, {Fonseca}, {Girma},
  {G{\" o}tting}, {Heinzelmann}, {Hermann}, {Heusler}, {Hofmann}, {Horns},
  {Wiedner}, {Wittek}, and {Remillard}]{spect_var_Mrk421_Aharonian02}
{Aharonian}~F, {Akhperjanian}~A, {Beilicke}~M, et~al.
\newblock {Variations of the TeV energy spectrum at different flux levels of
  Mkn 421 observed with the HEGRA system of Cherenkov telescopes}.
\newblock {\em \aap}, 393:\penalty0 89--99, Oct. 2002{\natexlab{b}}.

\bibitem[{Aharonian} et~al.(2003){Aharonian}, {Akhperjanian}, {Beilicke},
  {Bernl{\" o}hr}, {B{\" o}rst}, {Bojahr}, {Bolz}, {Coarasa}, {Contreras},
  {Cortina}, {Denninghoff}, {Fonseca}, {Girma}, {G{\" o}tting}, {Heinzelmann},
  {Hermann}, {Heusler}, {Hofmann}, et~al.]{1es1959Aharonian03}
{Aharonian}~F, {Akhperjanian}~A, {Beilicke}~M, et~al.
\newblock {Detection of TeV gamma-rays from the BL Lac 1ES 1959+650 in its low
  states and during a major outburst in 2002}.
\newblock {\em \aap}, 406:\penalty0 L9--L13, July 2003.

\bibitem[{Aharonian} et~al.(2005){Aharonian}, {Akhperjanian}, {Aye},
  {Bazer-Bachi}, {Beilicke}, {Benbow}, {Berge}, {Berghaus}, {Bernl{\" o}hr},
  {Bolz}, {Visser}, {V{\" o}lk}, and {Wagner}]{PKS2155_Aharonian05}
{Aharonian}~F, {Akhperjanian}~AG, {Aye}~KM, et~al.
\newblock {H.E.S.S. observations of PKS 2155-304}.
\newblock {\em \aap}, 430:\penalty0 865--875, Feb. 2005.

\bibitem[{Aharonian} et~al.(2000){Aharonian}, {Akhperjanian}, {Barrio},
  {Bernl{\" o}hr}, {Bojahr}, {Calle}, {Contreras}, {Cortina}, {Daum},
  {Deckers}, {Denninghoff}, {Fonseca}, {Gonzalez}, {Heinzelmann}, {Hemberger},
  {Hermann}, {He{\ss}}, {Heusler}, {Hofmann}, {Hohl}, {Horns}, {Ibarra},
  {Kankanyan}, {Kestel}, {Kettler}, {K{\" o}hler}, {Konopelko}, {Kornmeyer},
  {Kranich}, {Krawczynski}, {Lampeitl}, {Lindner}, {Lorenz}, {Magnussen},
  {Mang}, {Meyer}, {Mirzoyan}, {Moralejo}, {Padilla}, {Panter}, {Petry},
  {Plaga}, {Plyasheshnikov}, {Prahl}, {P{\" u}hlhofer}, {Rauterberg},
  {Renault}, {Rhode}, {R{\" o}hring}, {Sahakian}, {Samorski}, {Schmele},
  {Schilling}, {Schr{\" o}der}, {Stamm}, {V{\" o}lk}, {Wiebel-Sooth},
  {Wiedner}, {Willmer}, and {Wittek}]{2344Hegra2000}
{Aharonian}~FA, {Akhperjanian}~AG, {Barrio}~JA, et~al.
\newblock {HEGRA search for TeV emission from BL Lacertae objects}.
\newblock {\em \aap}, 353:\penalty0 847--852, Jan. 2000.

\bibitem[{Badran}(2001)]{2344Badran01}
{Badran}~HM.
\newblock {Recent Observations of 1ES 2344+514 Using the Whipple Gamma-Ray
  Telescope}.
\newblock In {\em AIP Conf. Proc. 587: Gamma 2001: Gamma-Ray Astrophysics},
  pages 281--285, 2001.

\bibitem[{Barth} et~al.(2003){Barth}, {Ho}, and
  {Sargent}]{Barth03_Black_hole_masses}
{Barth}~AJ, {Ho}~LC, {Sargent}~WLW.
\newblock {The Black Hole Masses and Host Galaxies of BL Lacertae Objects}.
\newblock {\em \apj}, 583:\penalty0 134--144, Jan. 2003.

\bibitem[{Beilicke} et~al.(2004)]{M87_HESS_Beilicke04}
{Beilicke}~M et~al.
\newblock {Observations of the giant radio galaxy M87 at TeV energies with
  H.E.S.S.}
\newblock In {\em XXII Texas Symposium on Relativistic Astrophysics}, 2004.
\newblock astro-ph/0504395.

\bibitem[{Bradbury} et~al.(1997){Bradbury}, {Deckers}, {Petry}, {Konopelko},
  {Aharonian}, {Akhperjanian}, {Barrio}, {Beglarian}, {Beteta}, {Contreras},
  and {Wirth}]{Mrk501_Bradbury97}
{Bradbury}~SM, {Deckers}~T, {Petry}~D, et~al.
\newblock {Detection of {$\gamma$}-rays above 1.5TeV from MKN 501.}
\newblock {\em \aap}, 320:\penalty0 L5--L8, Apr. 1997.

\bibitem[{Bussons-Gordo}(1998{\natexlab{a}})]{thesisBussons98}
{Bussons-Gordo}~J.
\newblock {\em {Derivation of the TeV Gamma-Ray Spectrum of Three Active
  Galaxies: Mrk 421, Mrk 501, and 1ES 2344+514}}.
\newblock PhD thesis, University College Dublin, July 1998{\natexlab{a}}.

\bibitem[{Bussons-Gordo}(1998{\natexlab{b}})]{2344Bussons98}
{Bussons-Gordo}~J.
\newblock {The TeV Energy Spectrum of the Active Galaxies 1ES 2344+514 and
  Markarian 501}.
\newblock In {\em Proceedings 16th ECRS, July 20-24}, pages 379+,
  1998{\natexlab{b}}.

\bibitem[{Catanese} et~al.(1998){Catanese}, {Akerlof}, {Badran}, {Biller},
  {Bond}, {Boyle}, {Bradbury}, {Buckley}, {Burdett}, {Bussons Gordo},
  {Carter-Lewis}, {Cawley}, {Connaughton}, {Fegan}, {Finley}, {Gaidos}, {Hall},
  {Hillas}, {Krennrich}, {Lamb}, {Lessard}, {Masterson}, {McEnery}, {Mohanty},
  {Quinn}, {Rodgers}, {Rose}, {Samuelson}, {Schubnell}, {Sembroski},
  {Srinivasan}, {Weekes}, {Wilson}, and {Zweerink}]{2344Catanese98}
{Catanese}~M, {Akerlof}~CW, {Badran}~HM, et~al.
\newblock {Discovery of Gamma-Ray Emission above 350 GeV from the BL Lacertae
  Object 1ES 2344+514}.
\newblock {\em \apj}, 501:\penalty0 616--623, July 1998.

\bibitem[{Catanese} et~al.(1997){Catanese}, {Boyle}, {Burdett}, {Bussons
  Gordo}, {Buckley}, {Carter-Lewis}, {Cawley}, {Fegan}, {Finley}, {Gaidos},
  {Hillas}, {Krennrich}, {Lamb}, {Lessard}, {Masterson}, {McEnery}, {Mohanty},
  {Quinn}, {Rodgers}, {Rose}, {Samuelson}, {Sembroski}, {Srinivasan}, {Weekes},
  and {Zweerink}]{2344Catanese1997ICRC}
{Catanese}~M, {Boyle}~PJ, {Burdett}~AM, et~al.
\newblock {First Results from a Search for TeV Emission from BL Lacs Out to Z =
  0.2.}
\newblock In {\em Proceedings of the 25th ICRC}, volume~3, pages 277--280,
  1997.

\bibitem[{Cawley}(1993)]{padding_Cawley93}
{Cawley}~MF.
\newblock {The Application of Noise Padding to the Cherenkov Imaging
  Technique}.
\newblock In {\em Towards a Major Cerenkov Detector}, pages 176--181. Universal
  Academy Press, 1993.

\bibitem[{Cawley} et~al.(1990){Cawley}, {Fegan}, {Harris}, {Kwok}, {Hillas},
  {Lamb}, {Lang}, {Lewis}, {Macomb}, {Reynolds}, {Schmid}, {Vacanti}, and
  {Weekes}]{detector_Cawley90}
{Cawley}~MF, {Fegan}~DJ, {Harris}~K, et~al.
\newblock {A high resolution imaging detector for TeV gamma-ray astronomy}.
\newblock {\em Experimental Astronomy}, 1:\penalty0 173--193, 1990.

\bibitem[{Chadwick} et~al.(1999){Chadwick}, {Lyons}, {McComb}, {Orford},
  {Osborne}, {Rayner}, {Shaw}, {Turver}, and {Wieczorek}]{pks2155_chadwick99}
{Chadwick}~PM, {Lyons}~K, {McComb}~TJL, et~al.
\newblock {Very High Energy Gamma Rays from PKS 2155-304}.
\newblock {\em \apj}, 513:\penalty0 161--167, Mar. 1999.

\bibitem[{Djannati-Atai } et~al.(1999){Djannati-Atai }, {Piron}, {Barrau},
  {Iacoucci}, {Punch}, {Tavernet}, {Bazer-Bachi}, {Cabot}, {Chounet},
  {Debiais}, {Degrange}, {Dezalay}, {Dumora}, {Espigat}, {Fabre}, {Fleury},
  {Fontaine}, {Ghesqui{\` e}re}, {Goret}, {Gouiffes}, {Grenier}, {Le Bohec},
  {Malet}, {Meynadier}, {Mohanty}, {Nuss}, {Par{\' e}}, {Qu{\' e}bert},
  {Ragan}, {Renault}, {Rivoal}, {Rob}, {Schahmaneche}, and
  {Smith}]{Mrk501_Djannati99}
{Djannati-Atai }~A, {Piron}~F, {Barrau}~A, et~al.
\newblock {Very High Energy Gamma-ray spectral properties of MKN 501 from CAT
  {\v C}erenkov telescope observations in 1997}.
\newblock {\em \aap}, 350:\penalty0 17--24, Oct. 1999.

\bibitem[{Douglas} et~al.(1996){Douglas}, {Bash}, {Bozyan}, {Torrence}, and
  {Wolfe}]{TexasRadioDouglas1996}
{Douglas}~JN, {Bash}~FN, {Bozyan}~FA, et~al.
\newblock {The Texas Survey of Radio Sources Covering -35.5 degrees {$<$}
  declination {$<$} 71.5 degrees at 365 MHz}.
\newblock {\em \aj}, 111:\penalty0 1945--1963, May 1996.

\bibitem[{Elvis} et~al.(1992){Elvis}, {Plummer}, {Schachter}, and
  {Fabbiano}]{EinsteinSlewSurvey_Elvis92}
{Elvis}~M, {Plummer}~D, {Schachter}~J, et~al.
\newblock {The Einstein Slew Survey}.
\newblock {\em \apjs}, 80:\penalty0 257--303, May 1992.

\bibitem[{Falomo} and {Kotilainen}(1999)]{optical_imaging_Falomo99}
{Falomo}~R, {Kotilainen}~JK.
\newblock {Optical imaging of the host galaxies of X-ray selected BL Lacertae
  objects}.
\newblock {\em \aap}, 352:\penalty0 85--102, Dec. 1999.

\bibitem[{Gaidos} et~al.(1996){Gaidos}, {Akerlof}, {Biller}, {Boyle},
  {Breslin}, {Buckley}, {Carter-Lewis}, {Catanese}, {Cawley}, {Fegan},
  {Finley}, {Hillas}, {Krennrich}, {Lamb}, {Lessard}, {McEnery}, {Mohanty},
  {Moriarty}, {Quinn}, {Rodgers}, {Rose}, {Samuelson}, {Schubnell},
  {Sembroski}, {Srinivasan}, {Weekes}, {Wilson}, and
  {Zweerink}]{Mrk421_Gaidos96}
{Gaidos}~JA, {Akerlof}~CW, {Biller}~SD, et~al.
\newblock {Extremely rapid bursts of TeV Photons from the active galaxy
  Markarian 421.}
\newblock {\em \nat}, 383:\penalty0 319--320, 1996.

\bibitem[{Giommi} et~al.(2000){Giommi}, {Padovani}, and
  {Perlman}]{exceptional_X-ray_var_Giommi_2000}
{Giommi}~P, {Padovani}~P, {Perlman}~E.
\newblock {Detection of exceptional X-ray spectral variability in the TeV BL
  Lac 1ES 2344+514}.
\newblock {\em \mnras}, 317:\penalty0 743--749, Oct. 2000.

\bibitem[{Gregory} and {Condon}(1991)]{87GBcatalogGreenbank_4.85GHz1991}
{Gregory}~PC, {Condon}~JJ.
\newblock {The 87GB catalog of radio sources covering delta between O and + 75
  deg at 4.85 GHz}.
\newblock {\em \apjs}, 75:\penalty0 1011--1291, Apr. 1991.

\bibitem[{Hartman} et~al.(1999){Hartman}, {Bertsch}, {Bloom}, {Chen},
  {Deines-Jones}, {Esposito}, {Fichtel}, {Friedlander}, {Hunter}, {McDonald},
  {Sreekumar}, {Thompson}, {Jones}, {Lin}, {Michelson}, {Nolan}, {Tompkins},
  {Kanbach}, {Mayer-Hasselwander}, {M{\" u}cke}, {Pohl}, {Reimer}, {Kniffen},
  {Schneid}, {von Montigny}, {Mukherjee}, and {Dingus}]{3rd_EGRET_Hartman99}
{Hartman}~RC, {Bertsch}~DL, {Bloom}~SD, et~al.
\newblock {The Third EGRET Catalog of High-Energy Gamma-Ray Sources}.
\newblock {\em \apjs}, 123:\penalty0 79--202, July 1999.

\bibitem[{Helene}(1983)]{upperLimit_Helene83}
{Helene}~O.
\newblock {Upper limit of peak area}.
\newblock {\em Nuclear Instruments and Methods in Physics Research A},
  212:\penalty0 319--322, Dec. 1983.

\bibitem[{Hillas}(1985)]{HillasParameters_Hillas1985}
{Hillas}~AM.
\newblock {Cerenkov light images of EAS produced by primary gamma}.
\newblock {\em NASA.~ Goddard Space Flight Center 19th Intern.~Cosmic Ray
  Conf., Vol.~3 p 445-448 (SEE N85-34862 23-93)}, 3:\penalty0 445--448, Aug.
  1985.

\bibitem[{Hillas} et~al.(1998){Hillas}, {Akerlof}, {Biller}, {Buckley},
  {Carter-Lewis}, {Catanese}, {Cawley}, {Fegan}, {Finley}, {Gaidos},
  {Krennrich}, {Lamb}, {Lang}, {Mohanty}, {Punch}, {Reynolds}, {Rodgers},
  {Rose}, {Rovero}, {Schubnell}, {Sembroski}, {Vacanti}, {Weekes}, {West}, and
  {Zweerink}]{CrabHillas98}
{Hillas}~AM, {Akerlof}~CW, {Biller}~SD, et~al.
\newblock {The Spectrum of TeV Gamma Rays from the Crab Nebula}.
\newblock {\em \apj}, 503:\penalty0 744--759, Aug. 1998.

\bibitem[{Holder} et~al.(2003){Holder}, {Bond}, {Boyle}, {Bradbury}, {Buckley},
  {Carter-Lewis}, {Cui}, {Dowdall}, {Duke}, {de la Calle Perez}, {Falcone},
  {Fegan}, {Fegan}, {Finley}, {Fortson}, {Gaidos}, {Gibbs}, {Gammell}, {Hall},
  {Hall}, {Hillas}, {Horan}, {Jordan}, {Kertzman}, {Kieda}, {Kildea}, {Knapp},
  {Kosack}, {Krawczynski}, {Krennrich}, {LeBohec}, {Linton}, {Lloyd-Evans},
  {Moriarty}, {M{\" u}ller}, {Nagai}, {Ong}, {Page}, {Pallassini}, {Petry},
  {Power-Mooney}, {Quinn}, {Rebillot}, {Reynolds}, {Rose}, {Schroedter},
  {Sembroski}, {Swordy}, {Vassiliev}, {Wakely}, {Walker}, and
  {Weekes}]{1es1959Holder03}
{Holder}~J, {Bond}~IH, {Boyle}~PJ, et~al.
\newblock {Detection of TeV Gamma Rays from the BL Lacertae Object 1ES 1959+650
  with the Whipple 10 Meter Telescope}.
\newblock {\em \apjl}, 583:\penalty0 L9--L12, Jan. 2003.

\bibitem[{Horan} et~al.(2002){Horan}, {Badran}, {Bond}, {Bradbury}, {Buckley},
  {Carson}, {Carter-Lewis}, {Catanese}, {Cui}, {Dunlea}, {Das}, {de la Calle
  Perez}, {D'Vali}, {Fegan}, {Fegan}, {Finley}, {Gaidos}, {Gibbs},
  {Gillanders}, {Hall}, {Hillas}, {Wakely}, and {Weekes}]{H1426_horan02}
{Horan}~D, {Badran}~HM, {Bond}~IH, et~al.
\newblock {Detection of the BL Lacertae Object H1426+428 at TeV Gamma-Ray
  Energies}.
\newblock {\em \apj}, 571:\penalty0 753--762, June 2002.

\bibitem[{Jarrett} et~al.(2003){Jarrett}, {Chester}, {Cutri}, {Schneider}, and
  {Huchra}]{2MASS_Jarrett03}
{Jarrett}~TH, {Chester}~T, {Cutri}~R, et~al.
\newblock {The 2MASS Large Galaxy Atlas}.
\newblock {\em \aj}, 125:\penalty0 525--554, Feb. 2003.

\bibitem[{Konopelko} et~al.(1999){Konopelko}, {Kettler}, and
  {HEGRA}]{2344Hegra99}
{Konopelko}~A, {Kettler}~J, {HEGRA}.
\newblock {TeV Gamma-Ray Observations of the BL Lac Object 1ES 2344+514 with
  the HEGRA System of Imaging Atmospheric Cherenkov Telescopes}.
\newblock In {\em Proceedings of the 26th ICRC}, volume~3, pages 426--429,
  1999.

\bibitem[{Krennrich} et~al.(2001){Krennrich}, {Badran}, {Bond}, {Bradbury},
  {Buckley}, {Carter-Lewis}, {Catanese}, {Cui}, {Dunlea}, {Das}, {de la Calle
  Perez}, {Fegan}, {Fegan}, {Finley}, {Gaidos}, {Gibbs}, {Gillanders}, {Hall},
  {Hillas}, {Holder}, {Horan}, {Jordan}, {Kertzman}, {Kieda}, {Kildea},
  {Knapp}, {Kosack}, {Lang}, {LeBohec}, {McKernan}, {Moriarty}, {M{\" u}ller},
  {Ong}, {Pallassini}, {Petry}, {Quinn}, {Reay}, {Reynolds}, {Rose},
  {Sembroski}, {Sidwell}, {Stanton}, {Swordy}, {Vassiliev}, {Wakely}, and
  {Weekes}]{Mrk421_Krennrich01}
{Krennrich}~F, {Badran}~HM, {Bond}~IH, et~al.
\newblock {Cutoff in the TeV Energy Spectrum of Markarian 421 during Strong
  Flares in 2001}.
\newblock {\em \apjl}, 560:\penalty0 L45--L48, Oct. 2001.

\bibitem[{Krennrich} et~al.(1999){Krennrich}, {Biller}, {Bond}, {Boyle},
  {Bradbury}, {Breslin}, {Buckley}, {Burdett}, {Gordo}, {Carter-Lewis},
  {Catanese}, {Cawley}, {Fegan}, {Finley}, {Gaidos}, {Hall}, {Hillas}, {Lamb},
  {Lessard}, {Masterson}, {McEnery}, {Mohanty}, {Moriarty}, {Quinn}, {Rodgers},
  {Rose}, {Samuelson}, {Sembroski}, {Srinivasan}, {Vassiliev}, and
  {Weekes}]{Mrk421Krennrich99}
{Krennrich}~F, {Biller}~SD, {Bond}~IH, et~al.
\newblock {Measurement of the Multi-TEV Gamma-Ray Flare Spectra of Markarian
  421 and Markarian 501}.
\newblock {\em \apj}, 511:\penalty0 149--156, Jan. 1999.

\bibitem[{Krennrich} et~al.(2002){Krennrich}, {Bond}, {Bradbury}, {Buckley},
  {Carter-Lewis}, {Cui}, {de la Calle Perez}, {Fegan}, {Fegan}, {Finley},
  {Gaidos}, {Gibbs}, {Gillanders}, {Hall}, {Hillas}, {Holder}, {Horan},
  {Jordan}, {Kertzman}, {Kieda}, {Kildea}, {Knapp}, {Kosack}, {Lang},
  {LeBohec}, {Moriarty}, {M{\" u}ller}, {Ong}, {Pallassini}, {Petry}, {Quinn},
  {Reay}, {Reynolds}, {Rose}, {Sembroski}, {Sidwell}, {Stanton}, {Swordy},
  {Vassiliev}, {Wakely}, and {Weekes}]{spect_var_Mrk421_Krennrich02}
{Krennrich}~F, {Bond}~IH, {Bradbury}~SM, et~al.
\newblock {Discovery of Spectral Variability of Markarian 421 at TeV Energies}.
\newblock {\em \apjl}, 575:\penalty0 L9--L13, Aug. 2002.

\bibitem[{Krennrich} et~al.(2003)]{Mrk421_Krennrich03}
{Krennrich}~F et~al.
\newblock {Hourly Spectral Variability of Mrk 421}.
\newblock In {\em Proceedings of the 28th ICRC}, 2003.
\newblock URL \url{astro-ph/0305419}.

\bibitem[{Lebohec} and {Holder}(2003)]{throughput_Lebohec03}
{Lebohec}~S, {Holder}~J.
\newblock {The cosmic ray background as a tool for relative calibration of
  atmospheric Cherenkov telescopes}.
\newblock {\em Astroparticle Physics}, 19:\penalty0 221--233, May 2003.

\bibitem[{Mohanty} et~al.(1998){Mohanty}, {Biller}, {Carter-Lewis}, {Fegan},
  {Hillas}, {Lamb}, {Weekes}, {West}, and {Zweerink}]{Mohanty98}
{Mohanty}~G, {Biller}~S, {Carter-Lewis}~DA, et~al.
\newblock {Measurement of TeV gamma-ray spectra with the Cherenkov imaging
  technique}.
\newblock {\em Astroparticle Physics}, 9:\penalty0 15--43, June 1998.

\bibitem[{Nikishov}(1962)]{AbsorptionNikishov62}
{Nikishov}~AI.
\newblock {Absorption of High-Energy Photons in the Universe}.
\newblock {\em Soviet Physics Jetp}, 14:\penalty0 393--394, Feb. 1962.

\bibitem[{Nilsson} et~al.(1999){Nilsson}, {Pursimo}, {Takalo}, {Sillanp{\"
  a}{\" a}}, {Pietil{\" a}}, and {Heidt}]{2d_photometric_Nilsson99}
{Nilsson}~K, {Pursimo}~T, {Takalo}~LO, et~al.
\newblock {Two-dimensional Photometric Decomposition of the TeV BL Lacertae
  Objects Markarian 421, Markarian 501, and 1ES 2344+514}.
\newblock {\em \pasp}, 111:\penalty0 1223--1232, Oct. 1999.

\bibitem[{Nishiyama} et~al.(1999)]{1es1959Nishiyama99}
{Nishiyama}~T et~al.
\newblock {1ES1959+650}.
\newblock In {\em Proceedings of the 26th ICRC}, volume~3, pages 370--373,
  1999.

\bibitem[{Patnaik} et~al.(1992){Patnaik}, {Browne}, {Wilkinson}, and
  {Wrobel}]{VLA_8.4GHz_1992}
{Patnaik}~AR, {Browne}~IWA, {Wilkinson}~PN, et~al.
\newblock {Interferometer phase calibration sources. I - The region 35-75 deg}.
\newblock {\em \mnras}, 254:\penalty0 655--676, Feb. 1992.

\bibitem[{Perlman} et~al.(1996){Perlman}, {Stocke}, {Schachter}, {Elvis},
  {Ellingson}, {Urry}, {Potter}, {Impey}, and
  {Kolchinsky}]{BLLacsFromEinsteinSS_Perlman96}
{Perlman}~ES, {Stocke}~JT, {Schachter}~JF, et~al.
\newblock {The Einstein Slew Survey Sample of BL Lacertae Objects}.
\newblock {\em \apjs}, 104:\penalty0 251--285, June 1996.

\bibitem[{Petry} et~al.(2002){Petry}, {Bond}, {Bradbury}, {Buckley},
  {Carter-Lewis}, {Cui}, {Duke}, {de la Calle Perez}, {Falcone}, {Fegan},
  {Fegan}, {Finley}, {Gaidos}, {Gibbs}, {Gammell}, {Hall},
  et~al.]{spectrum_H1426_Petry02}
{Petry}~D, {Bond}~IH, {Bradbury}~SM, et~al.
\newblock {The TeV Spectrum of H1426+428}.
\newblock {\em \apj}, 580:\penalty0 104--109, Nov. 2002.

\bibitem[{Petry} et~al.(1996){Petry}, {Bradbury}, {Konopelko}, {Fernandez},
  {Aharonian}, {Akhperjanian}, {Belgarian}, {Beteta}, {Contreras}, {Cortina},
  et~al.]{Mrk421_Petry96}
{Petry}~D, {Bradbury}~SM, {Konopelko}~A, et~al.
\newblock {Detection of VHE {$\gamma$}-rays from MKN 421 with the HEGRA
  Cherenkov Telescopes.}
\newblock {\em \aap}, 311:\penalty0 L13--L16, July 1996.

\bibitem[{Punch} et~al.(1992){Punch}, {Akerlof}, {Cawley}, {Chantell}, {Fegan},
  {Fennell}, {Gaidos}, {Hagan}, {Hillas}, {Jiang}, {Kerrick}, {Lamb},
  {Lawrence}, {Lewis}, {Meyer}, {Mohanty}, {O'Flaherty}, {Reynolds}, {Rovero},
  {Schubnell}, {Sembroski}, {Weekes}, and {Wilson}]{Mrk421_Punch92}
{Punch}~M, {Akerlof}~CW, {Cawley}~MF, et~al.
\newblock {Detection of TeV photons from the active galaxy Markarian 421}.
\newblock {\em \nat}, 358:\penalty0 477--478, Aug. 1992.

\bibitem[{Quinn} et~al.(1996){Quinn}, {Akerlof}, {Biller}, {Buckley},
  {Carter-Lewis}, {Cawley}, {Catanese}, {Connaughton}, {Fegan}, {Finley},
  {Gaidos}, {Hillas}, et~al.]{Mrk501_Quinn96}
{Quinn}~J, {Akerlof}~CW, {Biller}~S, et~al.
\newblock {Detection of Gamma Rays with E greater than 300 GeV from Markarian
  501}.
\newblock {\em \apjl}, 456:\penalty0 L83--L86, Jan. 1996.

\bibitem[{Reynolds} et~al.(1993){Reynolds}, {Akerlof}, {Cawley}, {Chantell},
  {Fegan}, {Hillas}, {Lamb}, {Lang}, {Lawrence}, {Lewis}, {Macomb}, {Meyer},
  {Mohanty}, {O'Flaherty}, {Punch}, {Schubnell}, {Vacanti}, {Weekes}, and
  {Whitaker}]{Survey1988-91_Reynolds93}
{Reynolds}~PT, {Akerlof}~CW, {Cawley}~MF, et~al.
\newblock {Survey of candidate gamma-ray sources at TeV energies using a
  high-resolution Cerenkov imaging system - 1988-1991}.
\newblock {\em \apj}, 404:\penalty0 206--218, Feb. 1993.

\bibitem[{Schroedter}(2004)]{thesis_Schroedter04}
{Schroedter}~M.
\newblock {\em {The Very High Energy Gamma-Ray Spectra of AGN}}.
\newblock PhD thesis, University of Arizona, 2004.

\bibitem[{Schroedter}(2005)]{EBL_Schroedter05}
{Schroedter}~M.
\newblock {Upper Limits on the Extragalactic Background Light from the Very
  High Energy Gamma-Ray Spectra of Blazars}.
\newblock {\em \apj}, 2005.

\bibitem[{Stecker}(1999)]{Steepening_Stecker99}
{Stecker}~FW.
\newblock {Intergalactic extinction of high energy gamma-rays}.
\newblock {\em Astroparticle Physics}, 11:\penalty0 83--91, June 1999.

\bibitem[{Stevens} and {Gear}(1999)]{SCUBA1999}
{Stevens}~JA, {Gear}~WK.
\newblock {Variations in the broad-band spectra of BL Lac objects: millimetre
  observations of an X-ray-selected sample}.
\newblock {\em \mnras}, 307:\penalty0 403--412, Aug. 1999.

\bibitem[{Tluczykont} et~al.(2003){Tluczykont}, {G{\"o}tting}, {Heinzelmann},
  and {HEGRA}]{2344Hegra03}
{Tluczykont}~M, {G{\"o}tting}~N, {Heinzelmann}~G, et~al.
\newblock {Observations of 54 Active Galactic Nuclei with the HEGRA Cherenkov
  Telescopes}.
\newblock In {\em Proceedings of the 28th ICRC}, volume~5, pages 2547--2550,
  2003.

\bibitem[{Urry} et~al.(2000){Urry}, {Scarpa}, {O'Dowd}, {Falomo}, {Pesce}, and
  {Treves}]{HST2000}
{Urry}~CM, {Scarpa}~R, {O'Dowd}~M, et~al.
\newblock {The Hubble Space Telescope Survey of BL Lacertae Objects. II. Host
  Galaxies}.
\newblock {\em \apj}, 532:\penalty0 816--829, Apr. 2000.

\bibitem[{Weekes} et~al.(1989){Weekes}, {Cawley}, {Fegan}, {Gibbs}, {Hillas},
  {Kowk}, {Lamb}, {Lewis}, {Macomb}, {Porter}, {Reynolds}, and
  {Vacanti}]{Crab_Nebula_Weekes89}
{Weekes}~TC, {Cawley}~MF, {Fegan}~DJ, et~al.
\newblock {Observation of TeV gamma rays from the Crab nebula using the
  atmospheric Cerenkov imaging technique}.
\newblock {\em \apj}, 342:\penalty0 379--395, July 1989.

\bibitem[{Weekes} et~al.(1987){Weekes}, {Lamb}, and
  {Hillas}]{HERCULES_Weekes87}
{Weekes}~TC, {Lamb}~RC, {Hillas}~AM.
\newblock {HERCULES - A new instrument for TeV astronomy}.
\newblock In {\em NATO ASIC Proc. 199: Very High Energy Gamma Ray Astronomy},
  pages 235--242, 1987.

\bibitem[{White} and {Becker}(1992)]{2344Greenbank1.4GHzWhite1992}
{White}~RL, {Becker}~RH.
\newblock {A new catalog of 30,239 1.4 GHz sources}.
\newblock {\em \apjs}, 79:\penalty0 331--467, Apr. 1992.

\bibitem[{Xie} et~al.(2002){Xie}, {Zhou}, {Dai}, {Liang}, {Li}, {Bai}, {Xing},
  and {Liu}]{photometric_monitoring_Xie2002}
{Xie}~GZ, {Zhou}~SB, {Dai}~BZ, et~al.
\newblock {Photometric monitoring of 12 BL Lacertae objects}.
\newblock {\em \mnras}, 329:\penalty0 689--699, Feb. 2002.

\end{thebibliography}
\end{document}